\documentclass[]{mn}

\usepackage{graphics}
\usepackage{rotating} 
\usepackage{amsfonts}

\def\aj{AJ}      
\def\apj{ApJ}      
\def\aap{A\&A}      

\def\apjs{ApJS}  
\def\aaps{A\&AS}  
\def\araa{ARA\&A} 

\def\mnras{MNRAS}        
         
\def\nat{Nature}

\def\ueber#1#2{{\setbox0=\hbox{$#1$}%
  \setbox1=\hbox to\wd0{\hss$ #2$\hss}%
  \offinterlineskip
  \vbox{\box1\box0}}{}}

\def\lesssim{\,\lower 1mm \hbox{\ueber{\sim}{<}}\,}
\def\grsim{\,\lower 1mm \hbox{\ueber{\sim}{>}}\,}

\title{The white dwarf luminosity  function. I. Statistical errors and
       alternatives}

\author[Geijo et al.]{Enrique M. Geijo$^{1}$,   
                      Santiago Torres$^{1,2}$,
                      Jordi Isern$^{2,3}$ and 
                      Enrique Garc\'{\i}a--Berro$^{1,2}$\\  
                      $^1$Departament de F\'\i sica Aplicada,  
                      Escola Polit\`ecnica Superior de Castelldefels,
                      Universitat Polit\`ecnica de Catalunya, \\
                      Avda. del Canal Ol\'\i mpic s/n,
                      08860 Castelldefels, Spain\\ 
                      $^2$Institut d'Estudis Espacials de Catalunya, 
                      c/Gran Capit\`a 2--4, Edif. Nexus  104, 
                      08034 Barcelona, Spain\\ 
                      $^3$Institut de Ci\`encies de l'Espai, C.S.I.C.,
                      Campus UAB, Facultat  de Ci\`encies, Torre C-5, 
                      08193 Bellaterra, Spain\\ }
\begin{document}

\maketitle

\begin{abstract} 
The white dwarf luminosity function is an important tool for the study
of the  solar neighborhood, since  it allows the determination  of the
age of the  Galactic disk.  Over the years,  several methods have been
proposed to compute galaxy  luminosity functions, from the most simple
ones ---  counting sample  objects inside a  given volume ---  to very
sophisticated ones  --- like the C$^-$  method, the STY  method or the
Choloniewski    method,    among    others.    However,    only    the
$1/\mathcal{V}_{\rm max}$ method is  usually employed in computing the
white  dwarf  luminosity function  and  other  methods  have not  been
applied  so  far  to  the observational  sample  of  spectroscopically
identified white dwarfs --- in  sharp contrast with the situation when
galaxy  luminosity  functions are  derived  from  a  large variety  of
samples.  Moreover,  the statistical  significance of the  white dwarf
luminosity function has also  received little attention and a thorough
study  still remains  to be  done.  In  this paper  we study,  using a
controlled synthetic  sample of white  dwarfs generated using  a Monte
Carlo simulator,  which is the  statistical significance of  the white
dwarf luminosity function and which  are the expected biases.  We also
present a  comparison between  different estimators for  computing the
white dwarf luminosity  function. We find that for  sample sizes large
enough  the  $1/\mathcal{V}_{\rm  max}$  method  provides  a  reliable
characterization of the white dwarf luminosity function, provided that
the   input   sample   is   selected  carefully.   Particularly,   the
$1/\mathcal{V}_{\rm  max}$ method  recovers well  the position  of the
cut--off of the white  dwarf luminosity function. However, this method
turns  out to be  less robust  than the  Choloniewski method  when the
possible incompletenesses  of the sample  are taken into  account.  We
also  find  that the  Choloniewski  method  performs  better than  the
$1/\mathcal{V}_{\rm max}$ method in  estimating the overall density of
white dwarfs,  but misses  the exact location  of the cut--off  of the
white dwarf luminosity function.
\end{abstract}


\section{Introduction}

The white dwarf luminosity function  is perhaps one of the most useful
tools for deriving important properties of the solar neighborhood.  In
particular,  but not only,  the disk  white dwarf  luminosity function
carries valuable  information about the  age of the Galaxy  (Winget et
al.  1987; Garc\'\i a--Berro et al.  1988; Hernanz et al. 1994; Richer
et al.   2000) and of the  stellar formation rate (Noh  \& Scalo 1990;
D\'\i az--Pinto et  al.  1994; Isern et al. 1995;  Isern et al. 2001).
Additionally,  the luminosity  function  of white  dwarfs provides  an
independent  test of  the theory  of dense  plasmas (Segretain  et al.
1994;  Isern  et al.   1997).   Finally,  the  white dwarf  luminosity
function  directly  measures  the  current  death rate  of  low--  and
intermediate--mass stars in the  local disk.  Consequently, a reliable
determination of the observational  white dwarf luminosity function is
of the maximum interest.

Previous observational  efforts, like the Palomar  Green Survey (Green
et  al. 1986)  have  provided us  with  an invaluable  wealth of  good
quality data.   Moreover, ongoing projects like the  Sloan Digital Sky
Survey (York  et al.  2000; Stoughton  et al.  2002;  Abazajian et al.
2003, 2004), the 2 Micron All Sky Survey (Skrutskie et al. 1997; Cutri
et al. 2003), the SuperCosmos Sky Survey (Hambly et al. 2001a; Hambly,
et  al.  2001b;  Hambly,  Irwin  \& MacGillivray  2001),  the 2dF  QSO
Redshift Survey (Vennes  et al.  2002), the SPY  project (Pauli et al.
2003),   and   others  will   undoubtely   increase   the  sample   of
spectroscopically identified white dwarfs with reliable determinations
of parallaxes and proper motions,  which are essential for an accurate
determination of  the white dwarf  luminosity function.  Last  but not
least, future space  missions like {\sl Gaia} (Perryman  et al.  2001)
will  increase the  sample of  known white  dwarfs with  very accurate
astrometric determinations  even further  (Torres et al.   2005), thus
allowing a precise and reliable determination of the properties of the
disk white dwarf population.

Over the years, several methods have been used to determine luminosity
functions for all sort of objects, ranging from main sequence stars to
galaxies.  These include the most simple ones (counting stars inside a
given volume)  to very  sophisticated ones ---  like the  C$^-$ method
(Lynden-Bell  1971), the  STY method  (Sandage et  al.  1979)  and the
Choloniewski method  (Choloniewski 1986).  In spite of  the variety of
methods currently  used to  estimate galaxy luminosity  functions, the
$1/\mathcal{V}_{\rm max}$  method (Schmidt 1968) is  the most commonly
used method  for estimating white dwarf  luminosity functions, though,
to the  best of  our knowledge,  nobody has yet  assesed in  depth its
statistical reliability for such  a purpose.  More specifically, up to
now only two works have studied how good the $1/\mathcal{V}_{\rm max}$
method performs in estimating the white dwarf luminosity function.  In
particular, Wood \& Oswalt (1998)  demonstrated by using a Monte Carlo
simulator that the $1/\mathcal{V}_{\rm max}$ method for proper--motion
selected  samples  is a  good  density  estimator,  although it  shows
important statistical  fluctuations when  estimating the slope  of the
bright  end  of  the  white  dwarf  luminosity  function.   Later  on,
Garc\'{\i}a--Berro   et  al.   (1999),   using  another   Monte  Carlo
simulator, corroborated the previous  study and, moreover, showed that
the standard procedure used by the $1/\mathcal{V}_{\rm max}$ method to
asign error  bars severely  understimates the size  of the  real error
bars for  a typical  sample of 200  objects.  Additionally  these last
authors also showed  that there was a bias in the  derived ages of the
solar  neighborhood, consequence of  the binning  procedure.  However,
the  most apparent  conclusion of  both  papers is  that, in  general,
selection  effects or,  simply,  the inherent  characteristics of  the
sample under  consideration have a strong  effect on the  shape of the
estimated white  dwarf luminosity function, despite  using an unbiased
estimator, like the $1/\mathcal{V}_{\rm max}$ method.

In  this   paper  we  assess  the  statistical   significance  of  the
observational white  dwarf luminosity function. For such  a purpose we
will use a controlled synthetic  sample of white dwarfs generated with
our Monte Carlo simulator.  We  discuss in depth which are the typical
biases introduced  by the procedures  used to select the  sample. This
includes  both  the bias  in  retrieving  the  correct slope  for  the
monotonically increasing branch of the white dwarf luminosity function
and, most  importantly, the bias obtained when  retrieving the precise
location of the  observed cut--off. This last point  is of the maximum
interest, since  the observed drop--off of the  white dwarf luminosity
is  currently  one of  the  best estimators  used  to  date the  local
neighborhood. We also  present an independent estimate of  the size of
the error bars.  Finally we discuss the advantages and shortcomings of
the several methods that can be used to obtain the observational white
dwarf  luminosity function and  we present  a set  of recommendations.
The reader should take into account  that in the present paper we only
discuss  the  sampling  biases  and  do  not  take  into  account  the
measurement  errors. Clearly,  the effects  of the  measurement errors
will affect the sampling  biases and viceversa.  Moreover, the effects
of  the measurement  errors  could  be as  important  as the  sampling
biases, although  this remains still to  be studied. Such  an study is
under  preparation and  will  be published  elsewhere.   The paper  is
organized  as  follows.   In  section  2  we  describe  the  different
estimators   most  commonly  used   today  for   obtaining  luminosity
functions.   Section  3  is   devoted  to  describe  the  Monte  Carlo
simulations used to compare the different methods previously described
in \S 2, whereas in Section 4 we apply the different estimators to our
Monte  Carlo samples.   Finally in  Section 5  we summarize  our major
findings and we draw our conclusions.

\section{The most commonly used luminosity function estimators}

\subsection{Schmidt's estimator for proper motion and magnitude 
            selected samples}

This method,  also known as the $1/\mathcal{V}_{\rm  max}$ method, was
first  introduced by  Schmidt  (1968)  in the  studies  of the  quasar
population.   Later on  Schmidt (1975)  extended it  to  proper motion
selected samples and Felten (1976) made a generalization of the method
introducing the dependence on the  direction of the sample. This turns
out  to be  useful when  studying  stellar samples  because the  scale
height of the Galactic disk introduces some biases.

Consider  a  sample  of  stars  having a  lower  proper  motion  limit
$\mu_{\rm  l}$ and  faint apparent  magnitude limit  $m_{\rm  f}$, the
maximum distance for which an object can contribute to the sample is:

\begin{equation}
r_{\max}=\min\left[\pi^{-1}(\mu/\mu_{\rm l});
\pi^{-1}10^{0.2(m_{\rm f}-m)}\right]
\end{equation}

\noindent where  $\pi$ is  the stellar parallax,  $\mu$ is  the proper
motion,  and  $m$ the  apparent  magnitude.   If  the sample  is  only
complete to a certain upper  proper--motion limit $\mu_{\rm u}$ and to
a bright  magnitude limit  $m_{\rm b}$, then  there is also  a minimum
distance for which an object contributes to the luminosity function:

\begin{equation}
r_{\min}=\max\left[ \pi^{-1}(\mu/\mu_{\rm u});
\pi^{-1}10^{0.2(m_{\rm b}-m)}\right]
\end{equation}

\noindent Additionally,  if the sample  only covers a fraction  of the
sky, $\beta$, then the maximum volume  in which a  star can contribute
is:

\begin{equation}
\mathcal{V}_{\max}=\frac{4\pi}{3}\beta(r_{\max}^{3}-r_{\min}^{3})
\end{equation}

The  luminosity function, $\varphi_M$,  is then  built by  binning the
sample  in $i\in(1,N)$ magnitude  bins and  performing a  weighted sum
over the objects in each  magnitude bin, $N_i$.  The weight with which
every object contributes  to the sum depends on  the maximum volume in
which it  could be  detected given its  apparent magnitude  and proper
motion:

\begin{equation}
\varphi(M_i)={\displaystyle\sum\limits_{j=1}^{N_i}}
\frac{1}{\mathcal{V}^j_{\max}}
\end{equation}

Though this  estimator is based on heuristic  appreciations about {\sl
how a good estimator should be}, it has been shown that it is unbiased
(Felten 1976).  However, the fact  that the estimator is unbiased does
not guarantee a good estimate  of the luminosity function if the input
data is  not properly selected.   More specifically, the  input sample
should  be complete and,  additionally, Takeuchi  et al.   (2000) have
pointed out  that the  $1/\mathcal{V}_{\rm max}$ method  for magnitude
selected samples is seriously affected when the input data --- even if
complete --- is clustered or,  more generally speaking, when the input
data  are not  homogenously  scattered.  This,  in  turn, affects  the
derived   shape  of  the   luminosity  function.    Consequently,  the
$1/\mathcal{V}_{\rm max}$ method should  only be used when homogeneity
and  completitude of  the sample  under consideration  are guaranteed.
This,  of  course,  is  not  an   easy  task,  and  for  most  of  the
observational samples is an {\sl ``a priori''} assumption.

\subsection{Maximum Likelihood estimators based on the probability of 
            selection}

Maximum  likelihood  estimators  are   based  on  the  probability  of
selecting  a given  object  in a  sample  and have  been  shown to  be
insensitive to sample  inhomogenities.  Moreover, by definition, these
estimators are  unbiased and have minimum variance  for large samples.
Two  variants  have  been  already  developed.  The  first  one  is  a
parametric  estimator  (Sandage  et  al.   1979),  hereafter  the  STY
estimator,   whereas   the   corresponding   non--parametric   version
(Efstathiou et  al.  1988) is  called the StepWise  Maximum Likelihood
estimator (SWML).  Both estimators are designed for magnitude selected
samples  and  their  performances  when evaluating  galaxy  luminosity
functions  have  been thoroughly  tested  using  detailed Monte  Carlo
simulations (Willmer 1997; Takeuchi et al. 2000).

Following  Luri et  al.   (1996) we  define  the likelihood  function,
$\mathcal{L}$, as the product  of the probability distributions of the
variables  of interest,  under the  assumption  that all  of them  are
statistically independent:

\begin{equation}
\mathcal{L}(\theta)={\prod \limits_{k=1}^{n_{x}}}
\mathcal{D}(x_{k}|\theta)
\end{equation}

\noindent  where $x$  is a  random variable  with  probability density
$\mathcal{D}(x_k|\theta)$ depending  on  a set of  unknown  parameters
$\theta=(\theta  _{1},\theta_{2},\ldots,\theta_{n})$  and realizations
$(x_{1},\ldots,x_{n_x})$.  The value of  $\theta$ that  maximizes this
function is  the maximum  likelihood estimator, $\theta_{\rm  ML}$, of
the parameters.

Observational  selection   may  be  modelled  by   introducing  a  new
probability  density, $\mathcal{M}(x_k|\theta)$,  with the  help  of a
normalization constant, $\mathcal{C}_{M}^{-1}$, which depends upon the
maximization    variables,    and    of    a    selection    function,
$\mathcal{S}(x_k)$:

\begin{equation}
\mathcal{M}(x_k|\theta)=\mathcal{C}_{M}^{-1}\mathcal{D}(x_k|\theta
)\mathcal{S}(x_k)
\end{equation}

A typical  example of such a  selection function is  that obtained for
the  case of  a sample  which  is complete  up to  a certain  limiting
apparent  magnitude,  $m_{\rm  lim}$.   In  this  case  the  selection
function   is   simply   a   Heaviside   function,   $\mathcal{S}(m_k)
=\Theta(m_k-m_{\lim})$.  Writing down  the probability distribution as
the  product of  the densities  of the  variables of  interest  in our
sample ---  absolute magnitude,  $M$, parallax, $\pi$,  and tangential
velocity, $v_{\rm  tan}$ --- we get the structure of the  STY and SWML
estimators for magnitude selected samples:

\begin{equation}
\mathcal{M}(M_k|\theta)\propto\frac{\varphi(M_{k}|\theta)}
{\int\varphi(M|\theta)dM}
\end{equation}

\begin{equation}
\mathcal{L}(\theta)={\prod \limits_{k=1}^{n_{x}}}
\mathcal{M}(M_{k}|\theta)
\end{equation}

It  becomes  obvious  from  the  definition  of  likelihood  that  the
statistical  independence  of  the  variables makes  the  maximization
process  insensitive to  the  distribution of  velocities  and to  the
parallax  probability density.  Note, however,  that for  the  case in
which we have a mixture of  populations (thin and thick disk and halo,
for instance) the velocities and  the absolute magnitude are no longer
independent variables.   The SWML method is then  obtained by adopting
for $\varphi$ as a stepwise function:

\begin{equation}
\varphi(M)=\sum_{i=1}^{N}\hat\varphi_{i}W(M_{i}-M)
\end{equation}

\noindent where the window function $W(M_{i}-M),$ is defined by:

\begin{equation}
W(M_{i}-M)\equiv\left\{
\begin{array}
[l]{cl}  1   &  {\rm  if  \  \   }M_{i}-\frac{\Delta  M}{2}\leq  M\leq
M_{i}+\frac{\Delta M} {2}\\ 0 & {\rm otherwise}
\end{array}
\right.
\end{equation}

\noindent and  $\hat\varphi_i$ yields  the luminosity function  of the
corresponding magnitude bin.  On the other hand, the  STY estimator is
obtained  by adopting  for  $\varphi$ a  parametric  function as,  for
example,  a Schechter function  (Schechter 1976).  For the  case under
study  we have  modified the  Schechter function  to adapt  it  to the
characteristics of our problem:

\begin{equation}
\varphi(M)\propto
10^{0.4(M^\star-M)(A+1)}\exp(-10^{1.6(M^\star-M)(A+1)})
\end{equation}

\noindent where the parameters $A$  and $M^\star$ are related with the
slope  and  with the  position  of the  cut--off  of  the white  dwarf
luminosity function, respectively.  It  is worth noticing that this is
a good  characterization of the luminosity function  when the cut--off
is sharp.  The real white dwarf luminosity function does not show such
a sharp cut--off but, instead,  a tail extending to fainter magnitudes
is observationally found (Oswalt et  al. 1996). This is the reason why
this  method cannot  be  applied  to real  samples  but to  simplified
synthetic samples in which this tail  is not present (see \S 3 below).
In both cases, the likelihood can be maximized using standard methods.
The main drawback  of these methods is that they  can obtain the shape
of  the  luminosity function  but  do  not  provide the  normalization
factor.

\subsection{Maximum likelihood estimators based on poissonian statistics}

As an  alternative to the  estimators shown in the  previous sections,
there  exist  other  maximum  likelihood  estimators  that  build  the
likelihood using a different approach.  Taking as a premise that local
distribution of objects in some pair  of variables of the sample has a
poissonian distribution,  it is possible  to define a likelihood  as a
function of the parameter space.  The first of such methods, the C$^-$
method, was proposed  by Lynden-Bell (1971) and was  later improved by
Choloniewski  (1986).  The  Choloniewski  method uses  simple data  to
define a  probabilistic model and  then a new likelihood,  by dividing
the parameter space (magnitude and  parallax in our case) in cells and
assuming poissonian  statistics for each cell.  This  method allows to
estimate  both the  shape of  the  luminosity function  and the  total
density of objects simultaneously.

We  consider a sample  with a  total number  of $N_{\rm  obj}$ objects
having  absolute   magnitudes  $M_i$  and   parallaxes  $\pi_i$,  with
$i=1,\ldots,N_{\rm obj}$.   The method is restricted to  a solid angle
$\Omega=4\pi\beta$.   Moreover, the absolute  magnitude is  assumed to
fall  within $M\in\lbrack  M_{0},M_{A}]$ and  the parallax  also obeys
$\pi\in\lbrack\pi_{0},\pi_{B}]$.  This defines a volume:

\begin{equation}
V_{t}=\frac{\Omega}{3}(r_{B}^{3}-r_{0}^{3})
\end{equation}

\noindent  where  $r_0=1/\pi_0$ and $r_B=1/\pi_b$.  Consequently,  the
number density can be written as

\begin{equation}
n=\frac{N_{c}}{V_{t}}
\end{equation}

\noindent  and the  luminosity  function and  spatial  density of  the
sample are, respectively:

\begin{equation}
\int^{M_{A}}_{M_{0}} \varphi(M)\,dM=n
\end{equation}

\begin{equation}
\int V_{t}\varrho(x,y,z)\,dxdydz=N_{\rm obj}
\end{equation}

\noindent  where $\varrho$  is the  spatial density  of the  sample of
objects.  The key point of the  method is to assume that the number of
objects in every interval $dM\,dxdydz$ is given by a poissonian random
process and that, consequently, the probability of finding $k$ objects
in each box can be modelled using the following expression:

\begin{equation}
P_{k}=\exp(-\lambda)\frac{\lambda^{k}}{k!}
\end{equation}

\begin{figure}
\vspace{6.0cm}
\includegraphics{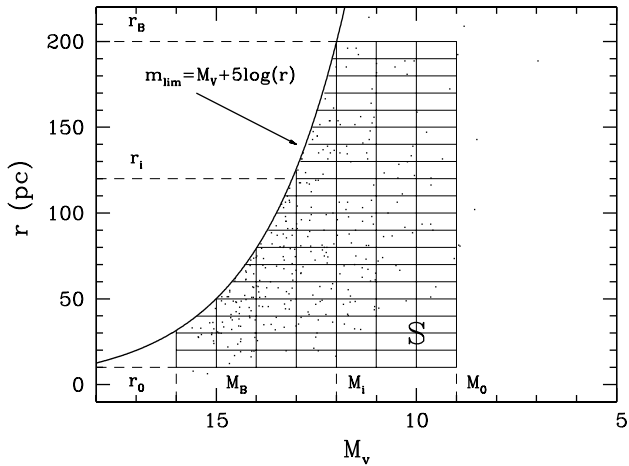} 
\caption{Distance versus absolute magnitude for a sample population of
         white dwarfs.  Also shown  are the limiting magnitude and the
         parameter  space, $S$,  adopted  for the  calculation of  the
         luminosity function  using the Choloniewski  method. See text
         for details.}
\label{Choloniewski}
\end{figure}

\noindent If, furthermore, the  distribution in absolute magnitude and
the spatial  density distribution are {\sl assumed}  to be independent
--- which may not  be the case if different  kinematic populations are
present in the sample --- we have:

\begin{equation}
\lambda=\frac{1}{n}\varphi(M)\varrho(x,y,z)\,dM\,dxdydz
\end{equation}

In  order  to  build  the  likelihood, the  previous  expressions  are
discretized  by  considering  the   distribution  of  objects  in  the
$(M,\pi)$ plane, and binning it  into square boxes (see figure 1).  We
set   $\Delta   M=\Delta\pi=\Delta$   and  $M_{i}=M_{0}+i\Delta$   and
$\pi_{j}=\pi_{0}+j\Delta$ with $j=0,\ldots,A$, and $i=0,\ldots,B$.  We
denote the number  of objects that can be found in  the box $(i,j)$ as
$N_{ij}$ and so the binned probability can be written as:

\begin{equation}
P_{N_{ij}}=\exp(-\lambda_{ij})\frac{\lambda_{ij}^{N_{ij}}}{N_{ij}!},
\end{equation}

\noindent with:

\begin{equation}
\lambda_{ij}=\frac{1}{n}\hat\varphi_{i}
\hat\varrho_{j}\Delta\frac{\Omega}{3}(r_{j} ^{3}-r_{j-1}^{3})
\end{equation}

\noindent  being  $\hat\varphi_i$ closely  related  to the  luminosity
function in the given magnitude interval through the expression:

\begin{equation}
\hat\varphi_{i}=\int^{M_{i-1}}_{M_{i}} \varphi(M)dM\times\left[
\int^{M_{i-1}}_{M_{i}} dM\right]^{-1},
\end{equation}

\noindent whereas the spatial density is given by:

\begin{equation}
\hat\varrho_{j}=\int_{r_{j-1}}^{r_{j}}\int_{\Omega}
\varrho(x,y,z)r^{2}\,dr\,d\Omega\times\left[\int_{r_{j-1}}^{r_{j}}
\int_{\Omega} r^{2}\,dr\,d\Omega\right]^{-1}
\end{equation}

Finally, we  mention that the Choloniewski likelihood must be computed
considering  the  selection  effects.  As  the  sample  only  provides
information for apparent magnitudes  up to a limiting magnitude $m\leq
m_{\rm lim}$, the  value of the number of objects in  each box is only
known for the region below the selection line. Thus, we have

\begin{equation}
\mathcal{L}(\theta)={\prod_{i=1}^{A}\prod_{j=1}^{B}}_{i,j\in S}
\exp(-\lambda_{ij})\frac{\lambda_{ij}^{N_{ij}}}{N_{ij}}
\end{equation}

\noindent where $S$ stands for grid in the parameter space (see figure
1). This likelihood can, again, be maximized using standard methods.

\section{The Monte Carlo Simulations}

Our Monte  Carlo simulator has been thouroughly  described in previous
papers  (Garc\'\i a--Berro  et  al.  1997;  Garc\'\i  a--Berro et  al.
2004) so here  we will only  summarize the most important  inputs.  We
have  used a  pseudo--random number  generator algorithm  (James 1990)
which  provides  a uniform  probability  density  within the  interval
$(0,1)$ and  ensures a  repetition period of  $\ga 10^{18}$,  which is
virtually   infinite  for   practical   simulations.   When   gaussian
probability functions are needed we have used the Box-Muller algorithm
(Press et al.  1986).

Since  we  want to  test  the  behaviour  of the  proposed  estimators
previously  discussed in  \S  2 under  different  assumptions for  the
underlying  white  dwarf  population  we  have analyzed  a  series  of
different scenarios  with controlled  stellar parameters.  In  a first
set of simulations  we have adopted the most  simple prescriptions for
the stellar  evolutionary inputs.   Specifically, we have  adopted the
most  simple cooling  law  (Mestel 1952).   Consequently, emission  of
neutrinos  was not considered.   Crystallization and  phase separation
were also  disregarded.  Additionally, for  all white dwarfs  we adopt
the same  cooling sequence, namely  that of a  typical $0.6\,M_{\sun}$
white dwarf, independently of  its respective mass.  Thus, the effects
of   the  mass   spectrum  of   white  dwarfs   are   also  completely
disregarded. The initial--to--final mass relationship for white dwarfs
and the main sequence lifetime  of their progenitors adopted here were
the analytical  expressions of Iben  \& Laughlin (1989).   Finally, no
bolometric corrections  were used.  Also a very  simple galactic model
has been used.  In particular, a standard initial mass function (Scalo
1998) and a constant volumetric star formation rate were adopted.  The
velocities have  been drawn from  normal laws taking into  account the
differential rotation of the disk and the peculiar velocity of the Sun
with respect to  the local standard of rest.  Since  the effect of the
spatial distribution  of the white dwarf population  can be important,
we have performed two kinds of simulations: in the first one a uniform
distribution was  used, whereas  for the second  one a  constant scale
height of 300~pc was assumed.

\begin{table} 
\begin{center}  
\caption{Summary of models}
\begin{tabular}{lccc}  
\hline
\hline 
  Model       & Cooling   & $r_{\rm max}$ & $z$-distribution\\ 
         & sequences &               & \\
\hline 
 1  & Mestel (1952)           &  250 pc & Uniform \\ 
 2  & Mestel (1952)           & 1800 pc & Uniform \\ 
 3  & Mestel (1952)           & 1800 pc & $h=300$~pc \\
 4  & Salaris et al. (2000)   & 1800 pc & Time-dependent $h$ \\ 
\hline
\hline
\end{tabular} 
\end{center} 
\end{table}

In a second  stage a set of more realistic  model simulations has also
been  performed.   For this  second  set  of  simulations the  cooling
sequences of Salaris et al. (2000) which incorporate the most accurate
physical  inputs  for   the  stellar  interior  (including  neutrinos,
crystallization and  phase separation) and reproduce the  blue turn at
low luminosities  (Hansen 1998) have  been used.  Also,  these cooling
sequences encompass the full range  of interest of white dwarf masses,
so a  complete coverage  of the  effects of the  mass spectrum  of the
white dwarf  population was taken  into account. Besides,  the spatial
density distribution is obtained from a scale height law (Isern et al.
1995) which  varies   with  time  and  is  related   to  the  velocity
distributions.

All the simulations  presented here are the average  of an ensemble of
400 independent  realizations. In all the  cases but in  the first one
the white  dwarf population  was modelled up  to distances  of $r_{\rm
max}=1800$~pc, in  order to avoid the  effects of the  border and were
normalized to the  local space density of white  dwarfs within 250~pc,
$n=0.5\times 10^{-3}$~pc$^{-3}$ for  $M_V < 12.75^{\rm mag}$ (Liebert,
Bergeron  \&   Holberg  2005).   For   model  1  we   adopted  $r_{\rm
max}=250$~pc  in  order to  test  the  effects  of a  distance-limited
sample.   Table   1  summarizes  the  main   characteristics  of  each
simulation.

\begin{figure}
\vspace{6.0cm}
\includegraphics{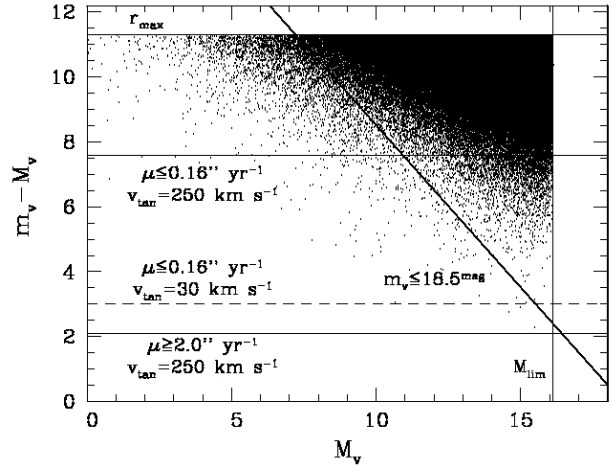}
\caption{Distance  modulus  versus  absolute  magnitude  for  a  whole
         population  of white  dwarfs within  a 1800  pc in  a uniform
         distribution.   Also  shown  are  the selection  criteria  in
         proper motion and apparent magnitude. See text for details.}
\label{selection effect}
\end{figure}

\section{Results}

\subsection{The $1/\cal{V}_{\rm max}$ method: slope, cut--off and binning}

We start  discussing model 1. Since  the final goal is  to compute the
white dwarf  luminosity function a  set of restrictions is  needed for
selecting  a subset  of white  dwarfs which,  in principle,  should be
representative of the whole white dwarf population. We have chosen the
following  criteria  for selecting  the  final  sample: $m_{\rm  V}\le
18.5^{\rm mag}$ and  $\mu\ge 0.16^{\prime\prime}\;{\rm yr}^{-1}$ as it
was done  in Oswalt et al.   (1996).  We do not  consider white dwarfs
with  very  small  parallaxes $(\pi\le  0.005^{\prime\prime})$,  since
these  are unlikely  to belong  to a  realistic  observational sample.
Additionally,  all white  dwarfs brighter  than $M_{\rm  V}\le 13^{\rm
mag}$ are included in the  sample, regardless of their proper motions,
since the  luminosity function of  hot white dwarfs has  been obtained
from  a catalog  of spectroscopically  identified white  dwarfs (Green
1980; Fleming et al.  1986)  which is assumed to be complete (Liebert,
Bergeron \& Holberg 2005).  Moreover, all white dwarfs with tangential
velocities larger than 250  km~s$^{-1}$ were discarded (Liebert et al.
1989) since these  would be probably classified as  halo members. With
all these  inputs the  white dwarf luminosity  function should  have a
constant slope and a very sharp cut--off, which depends on the adopted
age of the  disk.  Given the cooling law adopted  here the slope turns
out  to be 5/7  and, moreover,  considering that  the cut--off  of the
observational   white  dwarf   luminosity  function   is   located  at
$\log(L/L_{\sun})\simeq-4.6$ (Liebert et al.  1988) the adopted age of
the disk turns out to be 13~Gyr.

\begin{figure*}
\vspace{9.5cm}
\includegraphics{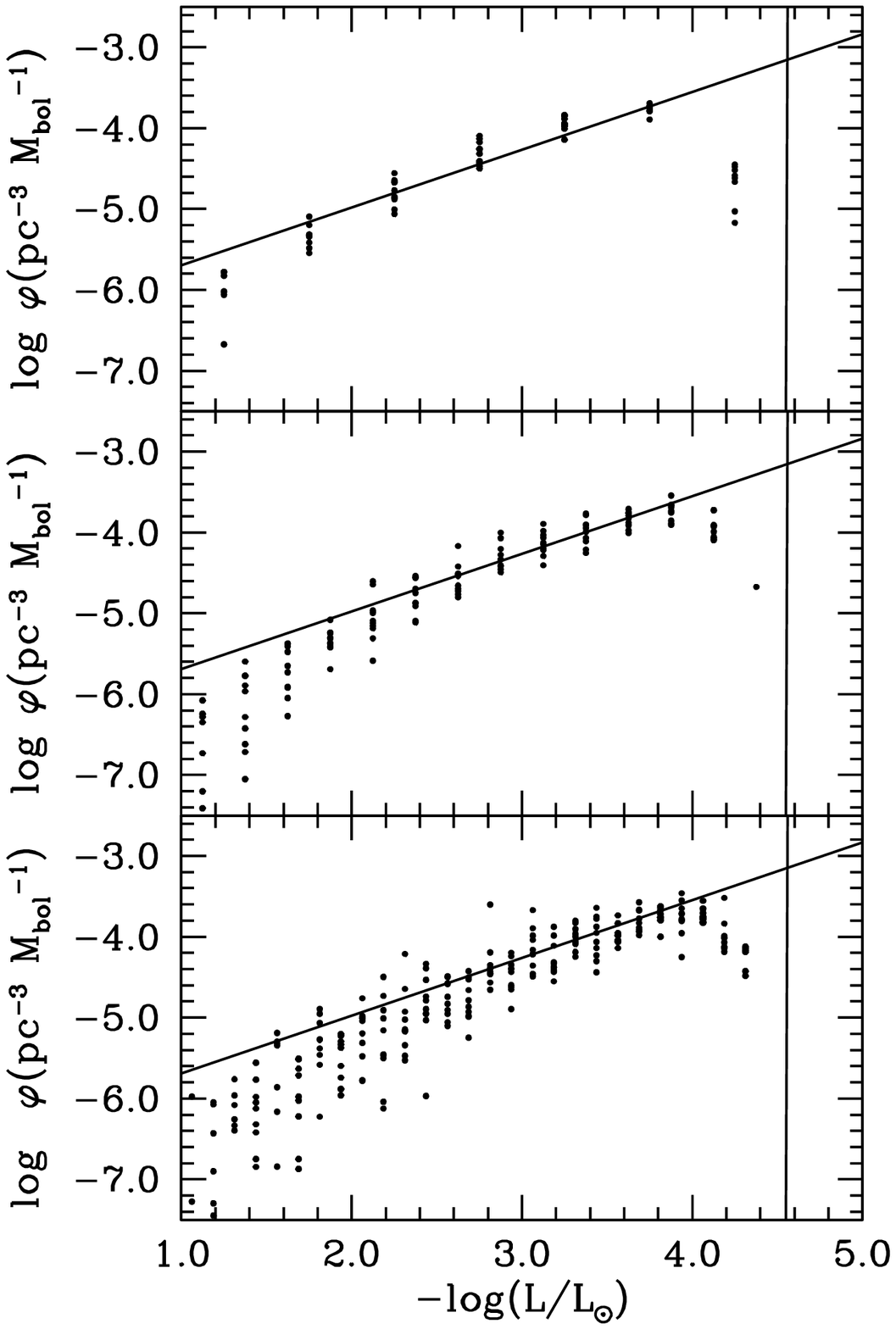}
\includegraphics{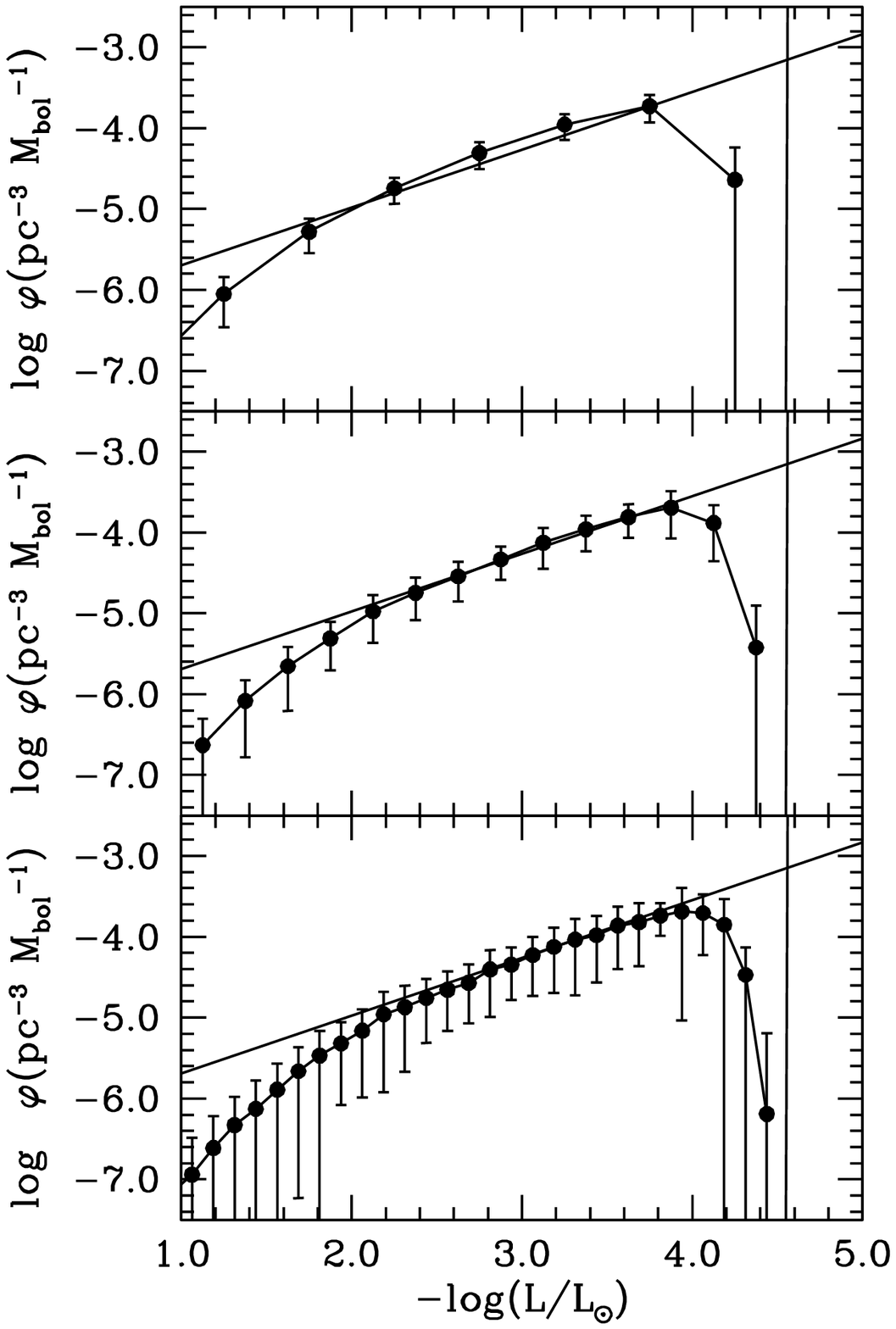}
\caption{{\sl{a)}} A sample of 20 realizations of the disk white dwarf
        luminosity function.  {\sl{b)}} Average and standard deviation
        of  the  400  realizations   of  the  white  dwarf  luminosity
        function.}
\label{realizations}
\end{figure*} 

In order to illustrate the effects of the previous restrictions in the
final sample, in  figure 2 we show the distance  modulus as a function
of the absolute $V$ magnitude  for the whole white dwarf population of
a  given  realization  of  our  Monte Carlo  simulations.   The  upper
horizontal  line corresponds  to  the maximum  distance  to which  our
synthetic population extends (1800 pc).  The vertical line corresponds
to the  limiting absolute magnitude, $M_{\rm lim}$,  which is directly
obtained from the  adopted age of the disk and  the Mestel cooling law
used in this set of  simulations. The diagonal line corresponds to the
limiting  magnitude  imposed  by  the  selection  criteria  previously
mentioned.  On the other hand, there is as well a maximum distance for
which a white dwarf could be  found within the proper motion limit. In
particular, an  object should have a tangential  velocity smaller than
250 km~s$^{-1}$ to  be considered as a disk  white dwarf, otherwise it
would be  considered a halo member.   Note, however, that  most of the
synthetic  white dwarfs  above  this line  have tangential  velocities
considerably  smaller  than  250~km~s$^{-1}$  and,  consequently  have
proper motions  smaller than the proper  motion cut. We  have drawn an
horizontal  line for this  maximum distance  ($r=330$~pc) for  which a
white dwarf  could be considered as  a disk member.   The upper dotted
line  in  this  diagram  corresponds  to the  distance  for  which  an
otherwise typical white  dwarf with $v_{\rm tan}=30$~km~s$^{-1}$ would
be included in the final  sample of white dwarfs ($r=65$~pc). Finally,
it  is worth  noticing  that the  currently  available proper  motions
surveys are  not sensitive to large proper  motions.  A representative
upper     cut    in     proper    motion     could     be    $\mu_{\rm
u}=2^{\prime\prime}$~yr$^{-1}$.  The  solid bottom horizontal  line in
Fig.   2 represents  the corresponding  distance for  a  high velocity
white dwarf with  $v_{\rm tan}=250$~km~s$^{-1}$, representative of the
halo population. As can be seen in Fig. 2 the effects of the selection
criteria are dramatical and only  an extremely small percentage of the
whole  white  dwarf  population  meets  the  selection  criteria  and,
consequently,  are  culled for  building  the  white dwarf  luminosity
function.

Fig.  3{\sl  a} shows 20  independent realizations of the  white dwarf
luminosity  function,  computed  with  the  $1/\mathcal{V}_{\rm  max}$
method, binned in 2 bins (top panel), 4 bins (middle panel) and 8 bins
(bottom  panel)  per  decade.   The  solid  lines  correspond  to  the
theoretical  expectations previously  described. That  is,  a straight
line with  constant slope  and a very  sharp cut--off at  the observed
position.  Each sample typically  contains about 300 white dwarfs, the
size of  the observational sample.   We recall that,  by construction,
our samples  are complete.   As can be  seen, there is  a considerable
spread  about the  theoretical expectations  and, moreover,  the white
dwarf  luminosity  function   is  underestimated  at  moderately  high
luminosities.  On the other hand, Fig.  3{\sl b} shows the average and
standard  deviation  of  the  400  realizations  of  the  white  dwarf
luminosity   function.   Again   it  is   clearly  visible   that  the
$1/\mathcal{V}_{\rm  max}$   method  considerably  underestimates  the
density of  white dwarfs with moderately high  luminosities.  In fact,
the $1/\mathcal{V}_{\rm max}$ method only recovers the right slope for
luminosities  smaller than  $\log(L/L_{\sun})\simeq  -2.2$.  Moreover,
the position of  the cut--off also depends on how  the data is binned,
its position being  more accurate for finer binning,  and it is always
located at  larger luminosities, a  direct consequence of  the binning
procedure, as already found by Garc\'\i a--Berro et al (1999).

\begin{figure*}
\vspace{9.5cm}
\includegraphics{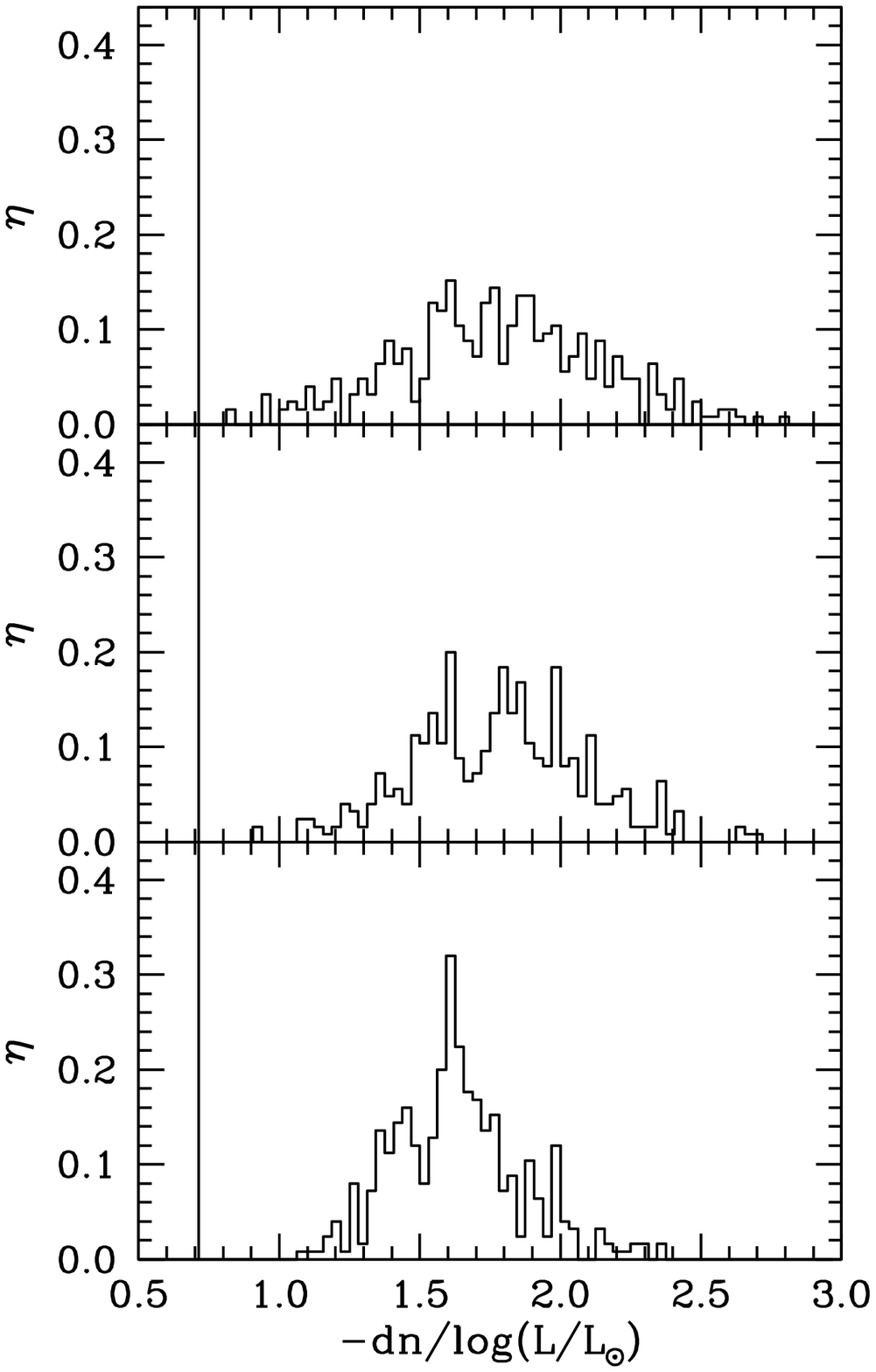}
\includegraphics{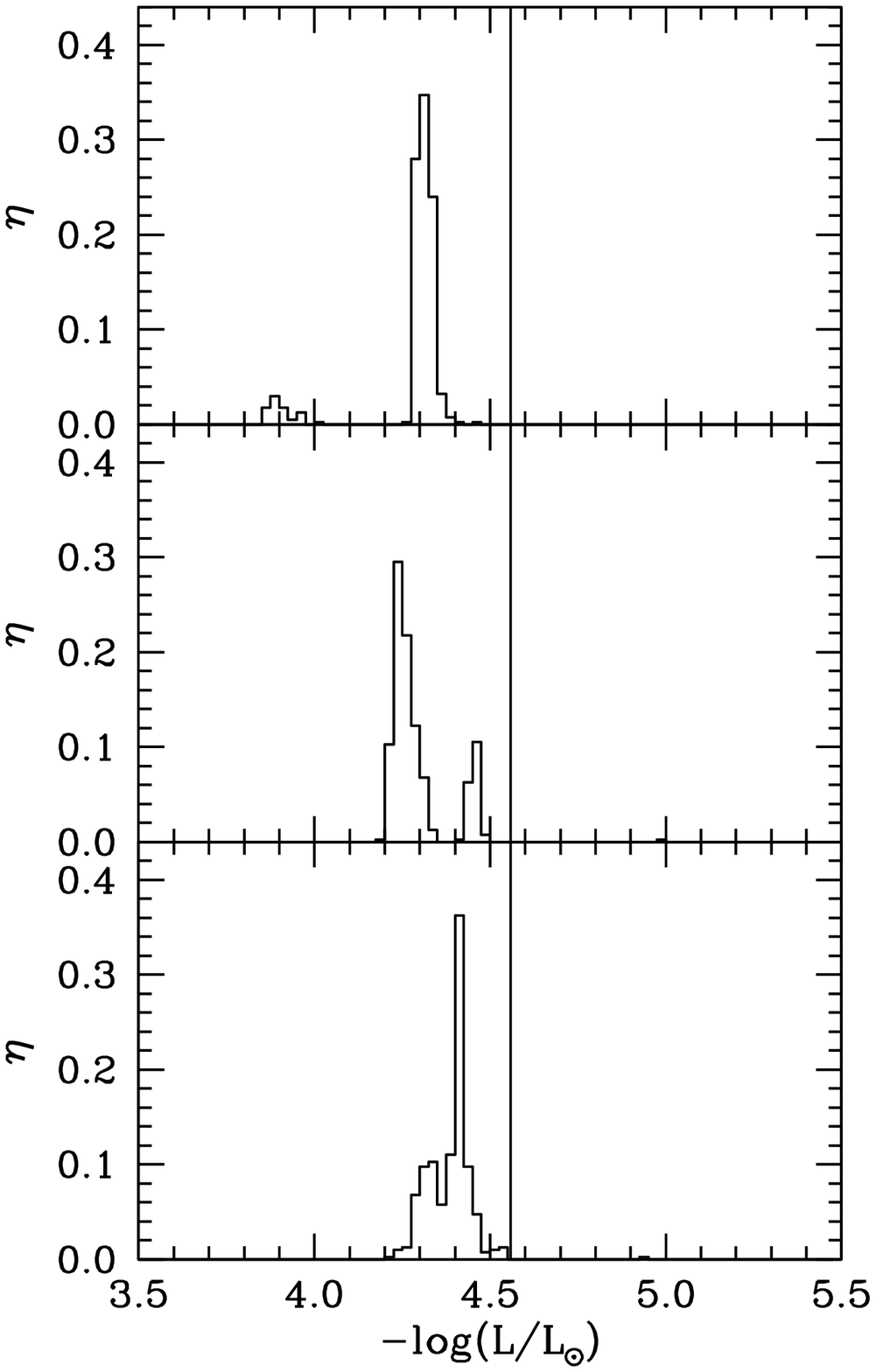}
\caption{{\sl{a)}}  Distribution   of  slopes  for  the  white dwarf
        luminosity  functions  of  Fig 3. {\sl{b)}} Distribution  of
        cut--offs  for  the  white  dwarf  luminosity  functions  of
        Fig. 3.}
\label{histograms}
\end{figure*} 

In order to quantify the previous  statements Fig. 4{\sl  a} shows the
frequency distribution of slopes  for the 400 independent realizations
of  the white  dwarf  luminosity function.   The  vertical solid  line
corresponds to the  theoretical value of the slope  of the white dwarf
luminosity  function  (5/7).   Obviously  {\sl all}  the  realizations
severely overestimate the slope and, moreover, there is a considerable
dispersion.  On  the other  hand, in Fig.~\ref{histograms}{\sl  b} the
distribution of  simulated cut--offs is shown.  Clearly, the finer the
binning  the  more  accurate  is  the determination  of the  cut--off.
However,  the dispersion  is relatively  small.   That is,  this is  a
systematic bias, which can be accounted for and ultimately corrected.

Another important  information that can  be readily obtained  from our
simulations is  how to compare  the observational procedure  to assign
error bars  --- basically assuming poissonian statistics  for each bin
(Liebert  et al.   1988) ---  with the  computed  standard deviations,
$\Delta \log \varphi_{\rm std}$.  To be more specific the contribution
of each star to the total  error budget in its luminosity bin, $\Delta
\log \varphi_{\rm  obs}$, is conservatively  estimated to be  the same
amount  that  contributes  to   the  resulting  density;  the  partial
contributions of each star in the  bin are squared and then added, the
final error is the square root of this value. Table 2 shows the result
of such a comparison. As can  be seen, we have found that, in general,
the standard  procedure to  assign error bars  severely underestimates
the observational error bars,  especially for the brigthest luminosity
bins of  the luminosity  function. Particularly, for  these luminosity
bins the error bars are underestimated by a factor of roughly $\approx
10$, whereas for the two  (more populated) dimmest luminosity bins the
error bar and the inherent statistical deviation are very similar.

\begin{table} 
\begin{center}  
\caption{A comparison of the error  bars of the white dwarf luminosity
         function computed using the $1/\mathcal{V}_{\rm max}$ method,
         $\Delta\log    \varphi_{\rm   obs}$,   with    the   standard
         observational  procedure  to   assign  error  bars  with  the
         intrinsic statistical  deviation of the  400 realizations for
         the  white   dwarf  population  of  model   1,  $\Delta  \log
         \varphi_{\rm std}$.  The  white dwarf luminosity function has
         been  obtained   by  binning  the  synthetic   data  in  four
         luminosity bins per decade.}
\begin{tabular}{ccc}  
\hline
\hline 
$-\log(L/L_{\sun})$  & $\Delta\log \varphi_{\rm obs}$  
& $\Delta \log \varphi_{\rm std}$\\ 
\hline 
1.625  & $-7.155$  & $-5.796$ \\ 
1.875  & $-7.222$  & $-5.535$ \\ 
2.125  & $-7.097$  & $-5.207$ \\
2.375  & $-5.886$  & $-5.016$ \\
2.625  & $-5.824$  & $-4.838$ \\
2.875  & $-6.000$  & $-4.689$ \\
3.125  & $-5.699$  & $-4.405$ \\
3.375  & $-5.398$  & $-4.291$ \\
3.625  & $-5.046$  & $-4.165$ \\
3.875  & $-5.155$  & $-3.925$ \\
4.125  & $-4.823$  & $-4.064$ \\
4.375  & $-4.921$  & $-5.063$ \\
\hline
\hline
\end{tabular} 
\end{center} 
\end{table}

\subsection{Looking for an alternative: other estimators}

\begin{figure*}
\vspace{9.5cm}
\includegraphics{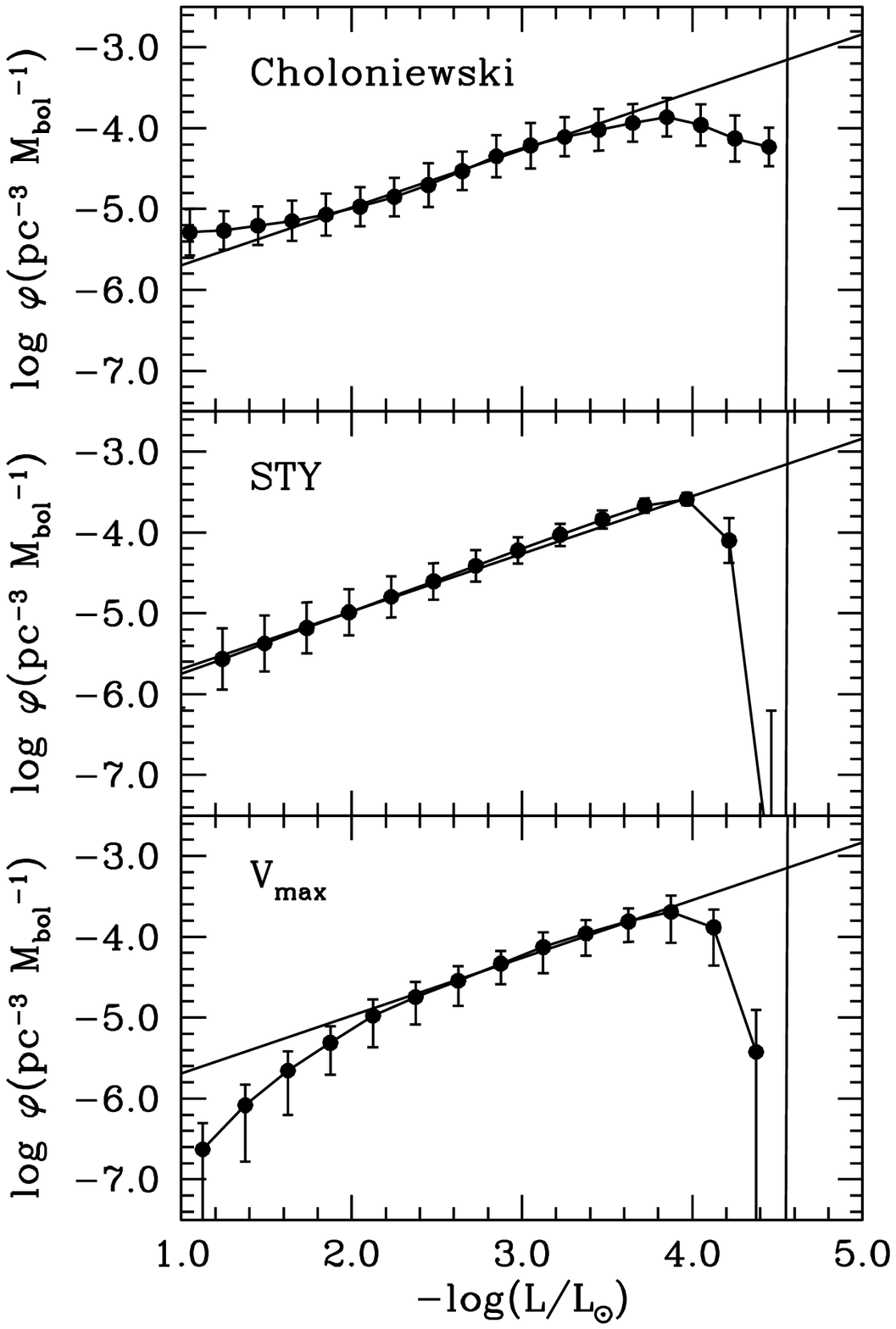}   
\includegraphics{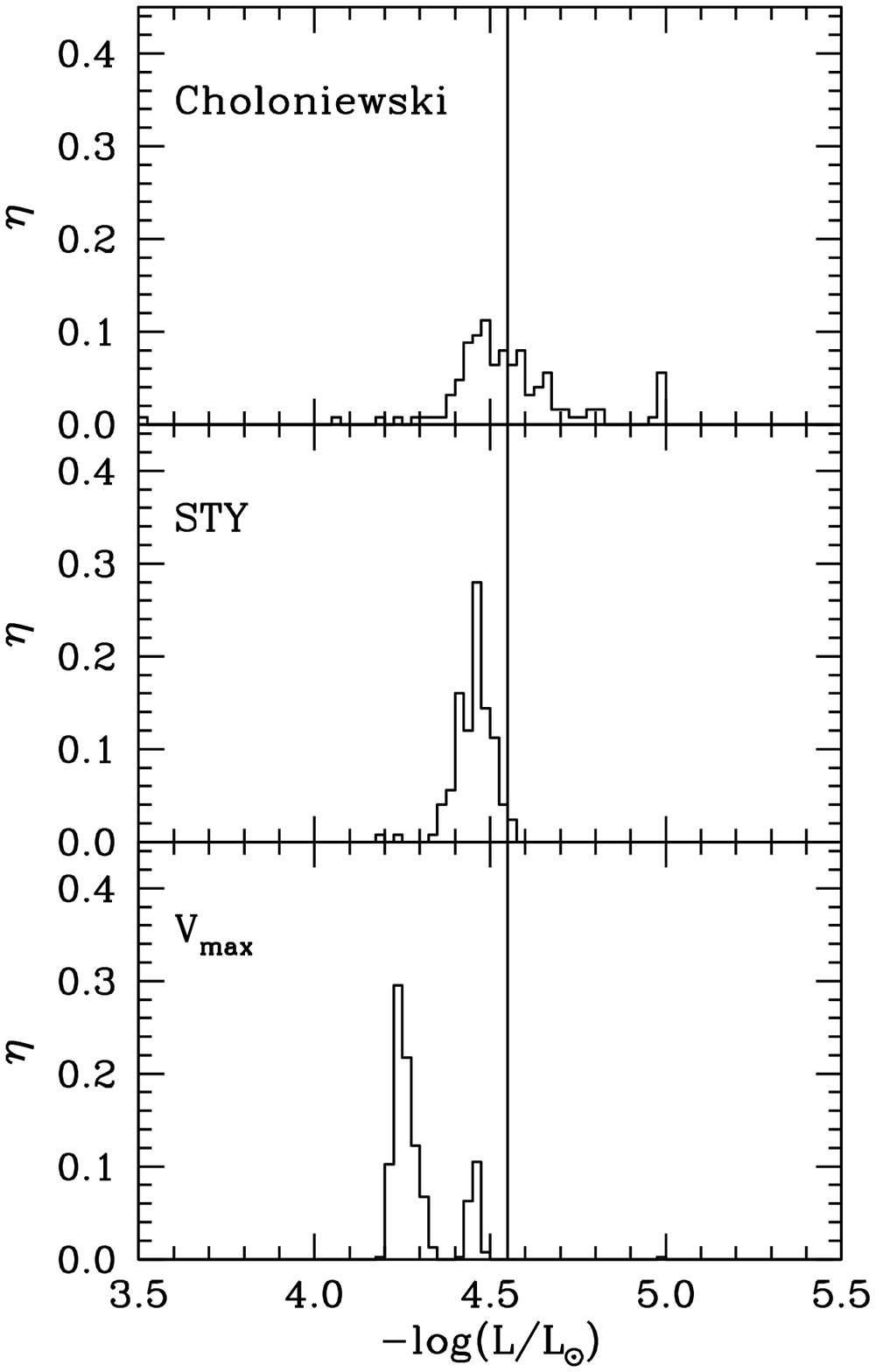} 
\caption{{\sl{a)}}  A   comparison  of  the   white  dwarf  luminosity
        functions   obtained  using   alternative   methods  and   the
        $1/\mathcal{V}_{\rm  max}$ method.  {\sl{b)}}  Distribution of
        cut--offs for the  different realizations  of the  white dwarf
        population using different methods.}
\label{alternative}
\end{figure*}

As  we  have shown,  the  $1/\mathcal{V}_{\rm  max}$  method does  not
provide satisfactory  answers with  regard to the  slope of  the white
dwarf  luminosity function  and the  position of the cut--off  for the
white dwarf populations  of model 1.  Thus, some  alternatives must be
explored.    As  previously  mentioned,   there  exist   several  such
alternatives. For  the sake of  conciseness here we will  discuss only
two  of  them:  the  STY   method  (Sandage  et  al.   1979)  and  the
Choloniewski method (Choloniewski  1986).  The SWML method (Efstathiou
et al.  1988)  gives results which are very similar  to the STY method
and, thus,  we will not  describe them in  detail for the  moment. All
three   methods  are   maximum  likelihood   methods  and   have  been
consistently used  to estimate galaxy luminosity  functions.  And this
is perhaps  their main  drawback since they  have not been  devised to
correct for the bias in proper motion. This is a characteristic of the
current white dwarf samples,  and the $1/\mathcal{V}_{\rm max}$ method
does correct  it. However,  as it will  be shown  below this is  not a
severe problem, at least for the  STY method.  We must recall that the
STY method provides  the shape of the luminosity  function but not its
normalization  (namely,  the true  density  of  objects), whereas  the
Choloniewski method provides both the shape of the luminosity function
and the density  of objects.  The selection criteria  used in this set
of simulations are  exactly the same used previously  for deriving the
white  dwarf luminosity function  using the  $1/\mathcal{V}_{\rm max}$
method.

Fig. 5{\sl a} shows a comparison of the results obtained for our model
1 using the different methods discussed here.  The bottom panel is the
same already shown in the middle  panel of Fig.  3{\sl b}.  The middle
panel shows the  results obtained using the STY method,  and as can be
seen, the STY  method recovers very well the  correct slope.  Finally,
the  top  panel shows  the  results  obtained  using the  Choloniewski
method.   Clearly,  this  method  underestimates  the  slope  at  high
luminosities.  The reason  for this is quite clear:  the statistics of
the brightest  luminosity bins are  not poissonian. In fact,  very few
white  dwarfs populate  these bright  luminosity bins  (note  that the
vertical scales in Fig.  3  and 5 are logarithmic).  Consequently, the
results  obtained  using the  Choloniewski  method  for the  brightest
luminosity bins are not correct.   However this method turns out to be
very useful since  we obtain the correct density  of white dwarfs.  In
all three  cases the error bars  are similar. On the  other hand, Fig.
5{\sl b}  shows the  position of the  cut--off for all  three methods.
The STY  method provides  better results than  the $1/\mathcal{V}_{\rm
max}$  method but,  undoubtely, the  Choloniewski method  performs the
best, although  with a larger  variance than the STY  method. Finally,
the statistical error  bars for all three methods  are rather similar,
being somehow smaller for the Choloniewski method.

\subsection{Extending the sample to larger distances}

\begin{figure}
\vspace{10cm}
\includegraphics{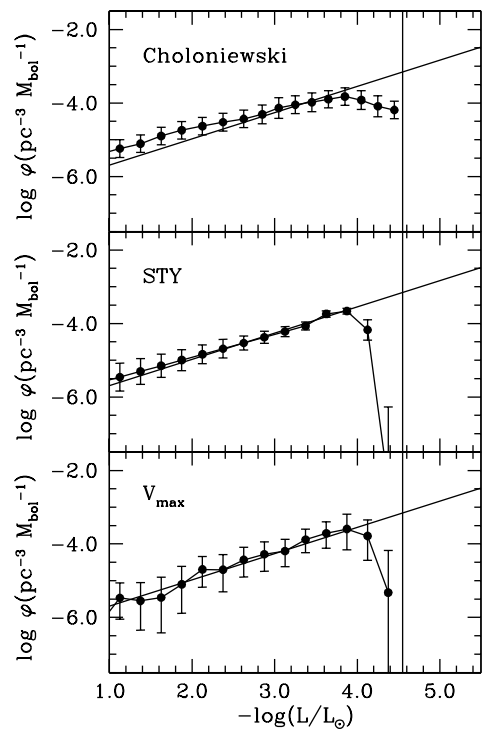}
\caption{Same  as  Fig. 5{\sl a}  for the  white  dwarf  population of 
         model 2.}
\end{figure}

One  possible reason  for the  systematic  bias found  when using  the
$1/\mathcal{V}_{\rm max}$ method to  obtain the white dwarf luminosity
function could be due to the  fact that bright objects can be found at
distances considerably larger than 250~pc, the maximum distance within
which  we have  distributed synthetic  white dwarfs  in model  1.  For
instance, an  object with  $\log(L/L_{\sun})=-1.0$ will be  within our
apparent  magnitude  selection criterion  even  if  it  is located  at
distances  as far as  1800~pc.  For  this reason  we have  applied the
different  luminosity function estimators  to a  sample with  a larger
maximum distance, in particular to  a sample of synthetic white dwarfs
distributed within  a sphere of radius  1800~pc (model 2  in Table 1).
Since  in  this sample  the  effects of  the  scale  height should  be
important we  have carried out  an additional simulation in  which the
synthetic white  dwarfs, instead of being distributed  according to an
uniform   density  law,   have  been   distributed  according   to  an
exponentially decrasing  density law. This model will  be discussed in
section 4.4 below. In both cases  the rest of the galactic and stellar
evolutionary  inputs  remain unchanged.  Note,  however,  that in  the
sample of spectroscopically identified  white dwarfs of McCook \& Sion
(1999) --- the primary observational source from which the white dwarf
luminosity function is  built --- the most distant  white dwarf with a
reliable parallax determination is located at $\sim 250$~pc.

In figure 6 the luminosity functions for the white dwarf population of
model 2 are  shown for the three estimators  previously discussed. The
results shown here are the  ensemble average and standard deviation of
a set of 20 realizations.  As can be seen, now the $1/\mathcal{V}_{\rm
max}$ estimator correctly matches the theoretical expectations for the
slope  of the  white dwarf  luminosity, although  with  a considerably
large variance for the brightest  luminosity bins. The reason for this
behavior will be discussed below, with the help of figure 7.  Also, it
is  interesting to  note  that for  this  set of  simulations the  STY
estimator  also yields  reasonable results,  whereas  the Choloniewski
estimator  largely  underestimates the  white  dwarf  density for  the
luminosity  bins  near  the  maximum  of the  white  dwarf  luminosity
function.   Finally, the  three  estimators obtain  the same cut--offs
previously obtained for model 1,  as it should be expected, given that
the population of  faint white dwarfs is drawn  from distances smaller
than 300 pc.

\begin{figure}
\vspace{6.0cm}
\includegraphics{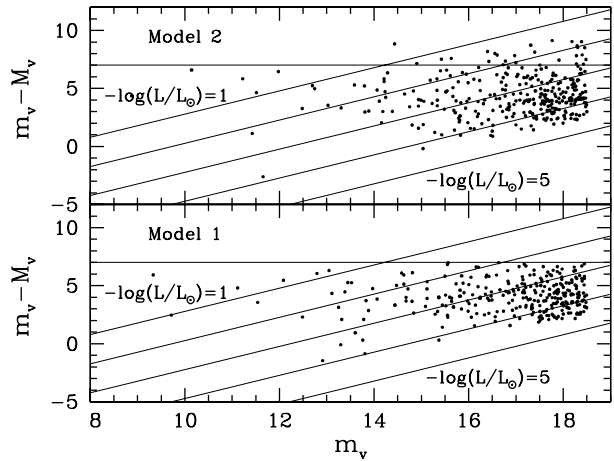} 
\caption{Distance modulus  versus magnitude for  a typical realization
         of  the  white dwarf  population  of  models  1 and  2.   The
         diagonal lines represent the  adopted bins of the white dwarf
         luminosity funcion.   The top panel  shows the results  for a
         sphere  of 1800~pc  with  a uniform  density  law (model  2),
         whereas the  bottom panel shows  the results for a  sphere of
         250~pc, also with a uniform density distribution.}
\end{figure}

In Fig.~7  the distribution of the  distance modulus as  a function of
the magnitude  for the synthetic  white dwarf populations of  models 1
(bottom panel) and 2 (top  panel) are shown.  We have also represented
the  diagonal lines  corresponding to  the adopted  bins of  the white
dwarf luminosity function,  from $\log(L/L_{\sun})=-1.0$ (top line) to
$-5.0$   (bottom   line).    The   horizontal  line   corresponds   to
$r=250$~pc. As can be seen, for model 2 (the sample distributed within
1800~pc) a  sizeable number of  intrinsically bright white  dwarfs (at
large  distances)  meet  the  selection  criteria  and,  consequently,
contribute  to the  white dwarf  luminosity function.   In  the sample
obtained  from model  1 these  intrinsically bright  white  dwarfs are
missing, and  this is the  reason why we  obtain a biased  white dwarf
luminosity  function.   It  is  nevertheless  worth  mentioning  three
important  points. Firstly, most  of the  spectroscopically identified
white dwarfs  in the catalog of  McCook \& Sion  (1999) have distances
smaller than  250~pc, as  can be seen  in top  panel of figure  8.  In
fact, only  three white dwarfs  in this catalog have  distances larger
than 250~pc and,  consequently, the slope of the  bright branch of the
white  dwarf luminosity function  obtained using  this catalog  as the
primary  source  of  observational   data  should  be,  in  principle,
questioned.  Secondly, the bright branch of the white dwarf luminosity
function  depends  primarily on  the  relative  strengths of  neutrino
leakage and  radiative losses.  Hence,  a robust determination  of the
slope  of the  bright branch  of the  white dwarf  luminosity function
turns out to be important  for deriving very useful constraints on the
physics of  cooling white dwarfs. And, finally,  and most importantly,
even in the  case of a distance limited sample, such  as that of model
1, the maximum likelihood  estimators and, specifically the STY method
and  the  Choloniewski method,  detect  the  deficit of  intrinsically
bright  white dwarfs  and, moreover,  they  are able  to retrieve  the
correct  slope.   Thus, these  estimators  are  much  more robust  and
provide more reliable white dwarf luminosity functions.

\subsection{The effects of the scale height}

\begin{figure}
\vspace{6.0cm}
\includegraphics{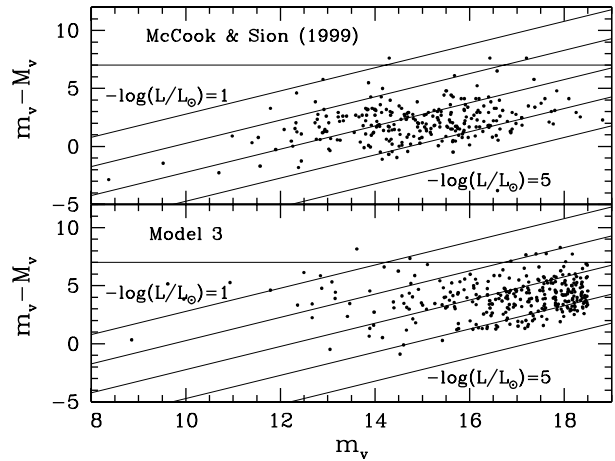} 
\caption{Same  as  figure  7  for  model  3  and  for  the  sample  of
         spectroscopically identified  white dwarfs of  McCook \& Sion
         (1999).}
\end{figure}

We  have also  tested which  would the  depedence of  the  white dwarf
luminosity function  on an exponentially decreasing  scale height law.
Hence,  instead of assuming  an uniform  distribution of  white dwarfs
within the computational volume, as it has been done so far for models
1 and 2, we have adopted a constant scale height of 300 pc (model 3 in
Table  1),  and  we   have  distributed  our  synthetic  white  dwarfs
accordingly  within a volume  of radius  1800 pc.   The results  for a
typical realization  of our Monte  Carlo simulations are shown  in the
bottom panel of  figure 8, where the distribution  of distance modulus
as a  function of the apparent  magnitude is shown  for this synthetic
population. As can be seen, for the brightest luminosity bin we obtain
now very few synthetic stars, as it should be expected, given that the
number  density  of  white   dwarfs  at  large  distances  is  heavily
suppressed  by  the  adopted  exponentially  decreasing  scale  height
law. In fact  we now obtain more  or less the same number  of stars in
both the synthetic sample and in the observational sample of McCook \&
Sion  (1999). Specifically,  for model  3 we  obtain on  average three
synthetic white dwarfs in the brightest luminosity bin, whereas in the
observational sample  of McCook  \& Sion (1999)  two white  dwarfs are
found in this  luminosity bin.  These numbers should  be compared with
the  corresponding one for  the population  of synthetic  white dwarfs
obtained  for model  2,  which,  on average,  is  13 synthetic  stars,
significantly  larger than  the previous  ones. However,  it  is worth
noticing  as well  that the  effects of  the scale  height and  of the
completeness  of  the  sample  under  study ---  especially  at  large
distances  ---  are  difficult   to  disentangle,  at  least  for  the
observational sample of McCook \&  Sion (1999). Clearly, a much better
observational  catalog, complete up  to distances  considerably larger
than  the scale  height of  Galactic disk,  should be  needed  to this
regard. For the  moment being, the possibility of  obtaining the value
of the scale height from a sample of tipicaly 300 white dwarfs remains
very remote.

\subsection{A realistic model}

As a final test we  have applied the several estimators discussed here
to  a more realistic  model of  the white  dwarf population,  which we
denote as model 4  (see Table 1).  We recall that for  model 4 we have
used the cooling sequences of  Salaris et al.  (2000), which encompass
the full range of masses of interest, as opposed to what has been done
up to  now, where the cooling  rate of any white  dwarf, regardless of
its  mass,  was   obtained  from  a  single  cooling   sequence  of  a
$0.6\,M_{\sun}$  white  dwarf.    Moreover,  these  cooling  sequences
incorporate  the  effects  of  neutrinos,  crystallization  and  phase
separation.   Consequently the  slope  of the  white dwarf  luminosity
function  is no longer  constant but,  instead, reflects  the relative
speed  of cooling  at a  given  luminosity. In  particular, for  those
luminosities where  neutrino cooling is  dominant the cooling  rate is
larger  and, consequently  the  slope of  the  white dwarf  luminosity
function turns out to be steeper, yielding {\sl less} white dwarfs for
these  luminosity  bins  when  compared  to  the  fiducial  luminosity
function  used up to  now.  Conversely,  for those  luminosities where
crystallization  and  phase   separation  are  the  relevant  physical
processes,   the   cooling   speed   is  smaller   (the   release   of
crystallization latent  heat and the gravitational  energy released by
phase separation  must be radiated away) and,  consequently, the slope
of the luminosity function is also steeper, yielding in this case {\sl
more}  white  dwarfs  for  these  luminosity bins  than  the  fiducial
luminosity function obtained from  Mestel's law (since they pile-up at
these luminosities  due to a  reduced cooling rate). For  this reason,
the expression  of Eq. (11) for  the STY estimator is  no longer valid
and, consequently, instead of using  the STY estimator for this set of
simulations  we   adopt  the  SWML  method,  which   provides  a  more
appropriate computational approach. We also  note that in this case we
have adopted  our full  model of Galactic  evolution, as  described in
detail in  Garc\'\i a--Berro  et al.  (1999).   Within this  model the
adopted scale--height depends on time --- being larger for past epochs
--- and, consequently,  since the adopted  star formation rate  in the
local  column has  been adopted  to  be constant  the volumetric  star
formation  rate  is  no   longer  constant.   Moreover,  the  velocity
dispersions  also  depend on  time  and,  thus,  the distributions  of
velocities are not perfectly gaussian as  it was the case for models 1
to 3.  As  a matter of fact our  Galactic evolutionary model naturally
incorporates the thin and the thick disk populations --- see Torres et
al. (2002).  However, the faint end of the disk white dwarf luminosity
function is  generally assumed  to be contaminated  by a yet  not well
known fraction of halo white  dwarfs (Reid 2005). Indeed, although the
peak  of the  halo white  dwarf luminosity  function is  located  at a
luminosity considerably fainter than that  of the cut--off of the disk
white dwarf  luminosity function (Isern  et al. 1998) some  halo white
dwarfs may be present in  faintest luminosity bins. This is the reason
why we apply  a very strict velocity cut  of 250~km~$s^{-1}$. While it
is true  that this  simple procedure does  not completely  remove high
velocity populations  it is also  true that the results  obtained with
model 4 represent a step in the right direction.  Finally we point out
that  in order  to keep  consistency with  the  simulations previously
described we have adopted the same  age of the disk. Since the cooling
sequences of  model 4 incorporate  the effects of  crystallization and
phase  separation, which  introduce a  sizeable delay  in  the cooling
times, the cut--off  in the white dwarf luminosity  functions moves to
fainter luminosities accordingly.

At this  point of  the discussion  of our results  it is  important to
realize that up  to now we have always had  a ``template'' white dwarf
luminosity function to  which we could compare. This  template was the
very  simple  luminosity function  already  shown  in  Figs.~3, 5  and
6. Given the stellar and galactic  inputs adopted for model 4, a white
dwarf luminosity function with a  perfectly constant slope and a sharp
cut--off is a poor  characterization of the  theoretical expectations.
However,  we  can easily  obtain  a  template  white dwarf  luminosity
function in the  following way.  We recall that,  by construction, our
samples are complete,  although we only select about  300 white dwarfs
using the  selection criteria already discussed  before.  However, our
simulations do provide the whole population of white dwarfs. Hence, we
can obtain the {\sl real} luminosity function by simply counting white
dwarfs in the computational volume.  This is done for all realizations
and then we obtain the average. The result is depicted as a solid line
in  Fig.   9,  where we  also  show  the  results obtained  using  the
Choloniewski method (upper panel),  the SWML method (middle panel) and
the  $1/\mathcal{V}_{\rm  max}$  method  (bottom  panel),  with  their
computed error  bars.  As can be  seen the cut--off of the white dwarf
luminosity  function has  moved to  fainter luminosities,  its precise
location being now $\log(L/L_{\sun})\simeq -4.8$, a direct consequence
of crystallization and phase separation.

\begin{figure}
\vspace{10cm}
\includegraphics{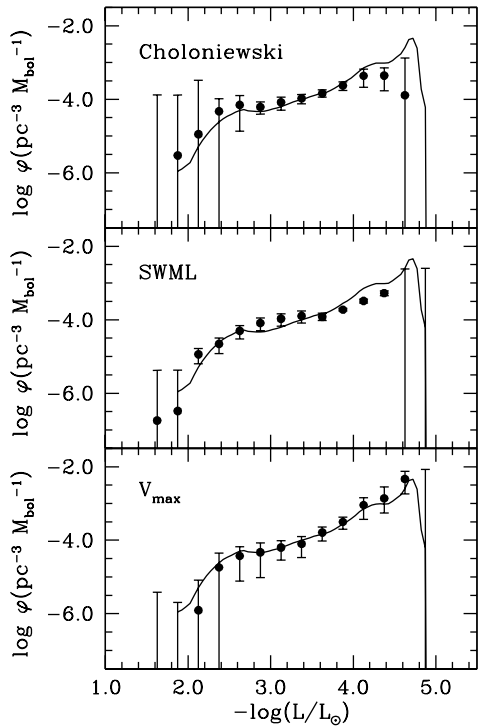}
\caption{White  dwarf  luminosity  function  for  model  4  using  the
         different  estimators   under  study  (dots),   our  template
         luminosity function is shown as a solid line.}
\end{figure}

Fig. 9  clearly shows that the performance  of the $1/\mathcal{V}_{\rm
max}$ method is  superb, since this method nicely  fits both the shape
of  the   white  dwarf  luminosity   function  and  the   position  of
cut--off. The  SWML method (middle panel  of Fig. 9)  also fits pretty
well the  shape of the white  dwarf luminosity function,  but the last
two (faint) luminosity bins  are poorly determined.  Consequently, the
determination of  the cut--off of the white  dwarf luminosity function
is subject to  a large variance, and individual  simulations can yield
very  different  results  for  the  age of  the  disk.   Finally,  the
Choloniewsky method (top panel  of Fig.  9) clearly underestimates the
number of faint  white dwarfs (the peak in  the white dwarf luminosity
function) and does  not reproduce the real cut--off. All in all, for a
sample  of about  300 white  dwarfs,  and when  all the  observational
biases are correctly taken  into account the $1/\mathcal{V}_{\rm max}$
performs best.

Also, some  computational details are worth mentioning.  The first one
is that the  computational load of the two  maximum likelihood methods
is much larger than that of the $1/\mathcal{V}_{\rm max}$ method. This
does not  pose a severe  problem when samples  with a small  number of
objects are analyzed  but it is a point to  be considered when samples
containing a large number of white dwarfs, like that of the SDSS which
will be the object of \S 4.6, are studied. The second important remark
is that for  a sample size of 300 white dwarfs  the convergence of the
two maximum likelihood  methods is slow, a consequence  of the minimum
being very shallow.

\subsection{The future: the SDSS}

Very  recently, a  sample  of  white dwarfs  selected  from the  Sloan
Digital Sky  Survey Data Release  3 (SDSS DR3) combined  with improved
proper  motions from the  USNO-B has  derived a  preliminary (although
very  much improved)  white dwarf  luminosity function  based  on 6000
stars (Harris et al.  2006). We emphasize at this point that we do not
aim to perform a full analysis  of the sample of Harris et al. (2006),
but a preliminary assessment of it. A detailed analysis of this sample
is out of the scope of this paper and we postpone it for a forthcoming
publication.   The  white  dwarf  luminosity  function  of  Harris  et
al. (2006) has been built  using the following selection criteria. The
survey area of  the SDSS is mostly centered  around the North Galactic
Cap  and covers  an area  of  $5282^{\circ^2}$.  For  our Monte  Carlo
simulations  we have  adopted the  precise  geometry of  the SDSS,  an
elliptical region  centered at $\alpha=  12^{\rm h}\, 20^{\rm  min}$ ,
$\delta= +32.8^\circ$, whose minor axis  is the meridian at that right
ascension, with extent $\pm  55^\circ$ in declination.  The major axis
is  the great circle  perpendicular to  that, and  the extent  is $\pm
65^\circ$; it  extends from about  $7^{\rm h}\, 6^{\rm min}$  to about
$17^{\rm h}\,  34^{\rm min}$. From  the original sample of  6000 stars
Harris   et    al.    (2006)    have   only   selected    stars   with
$\mu>20$~mas~yr$^{-1}$ and,  thus, we disregard all  white dwarfs with
proper motions  smaller than this value.  Additionally,  Harris et al.
(2006) use  the   reduced  proper  motion  $H_g=g+5\log\mu+5=M_g+5\log
V_{\rm tan}-3.379$,  where $g$ is the SDSS  magnitude, to discriminate
between main  sequence stars  and white dwarfs,  since the  latter are
typically  5--7 magnitudes less  luminous than  subdwarfs of  the same
color.  Moreover, they require that all white dwarfs must have $V_{\rm
tan}>30$~km/s to enter in the final sample, and this is what we adopt.
An  additional  criterion  is   that  all  white  dwarfs  should  have
$15.0<g<19.5$.  We  have selected only white  dwarfs with $15.0<m_{\rm
V}<19.5$. The final  size of the sample used to  built the white dwarf
luminosity function is of about $\sim 2000$ stars.

\begin{figure}
\vspace{10cm}
\includegraphics{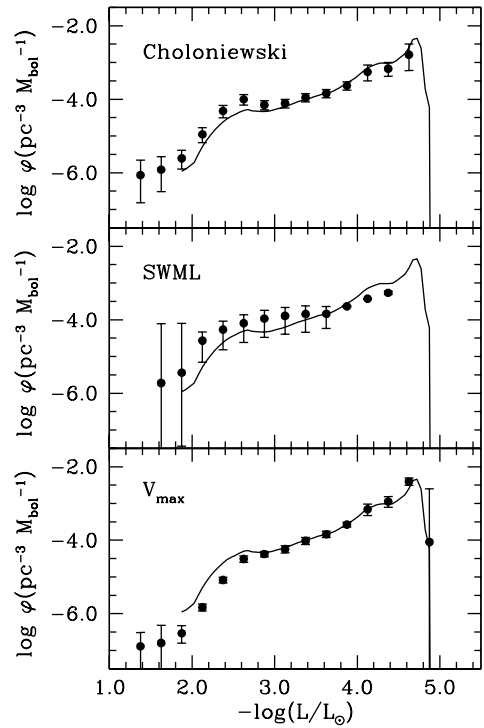}
\caption{Same as Fig.~9 for a  sample of 2000 white dwarfs, the sample
         size of the SDSS.}
\end{figure}

With  all  these  restrictions   we  have  computed  the  white  dwarf
luminosity function with the inputs  of model 4. The results are shown
in  Fig.~10.  As it  has been  done so  far, we  show the  white dwarf
luminosity function computed with the Choloniewski method (top panel),
the  SWML  method (middle  panel)  and  the $1/\mathcal{V}_{\rm  max}$
method (bottom  panel). Clearly, both the Choloniewsky  method and the
$1/\mathcal{V}_{\rm max}$  method perform very well,  whereas the SWML
method  misses  the  maximum  and  the  cut--off  of the  white  dwarf
luminosity  function and,  moreover,  the variance  for the  brightest
luminosity  bins  is   much  larger  than  those  of   the  other  two
methods. For the Choloniewski method  the last luminosity bin does not
show up, but it should be  taken into account that the the variance of
the  last  bin  of   the  $1/\mathcal{V}_{\rm  max}$  method  is  very
large. One comment  is in order regarding this last  method. As can be
seen in  Fig.~10, the $1/\mathcal{V}_{\rm  max}$ method underestimates
the  white  dwarf luminosity  function  for  the brightest  luminosity
bins. This  is a  consequence of the  adopted galactic inputs  for our
white dwarf  population and, more  specifically, of the  adopted scale
height.   Since we  are using  the original  $1/\mathcal{V}_{\rm max}$
method, whithout correcting for the  scale height, the number of white
dwarfs per  unit volume and  magnitude interval is  underestimated for
the brightest luminosity bins,  where the survey extends to relatively
large  distances.    On  the  other  hand,   the  Choloniewski  method
overestimates the white dwarf density for these luminosity bins.  Note
however,   that    for   the   intermediate    luminosity   bins   the
$1/\mathcal{V}_{\rm max}$  method matches very  well the shape  of the
white dwarf luminosity function.  All in all, except for the brightest
luminosity bins, the $1/\mathcal{V}_{\rm  max}$ method provides a very
good characterization of the white dwarf luminosity function. Finally,
and contrary  to what was found  in \S 4.1  for a sample of  300 white
dwarfs, the  observational procedure for  assigning the error  bars to
the white dwarf luminosity function is fair for a sample of 2000 white
dwarfs.

\subsection{The incompleteness of the sample}

\begin{figure}
\vspace{10cm}
\includegraphics{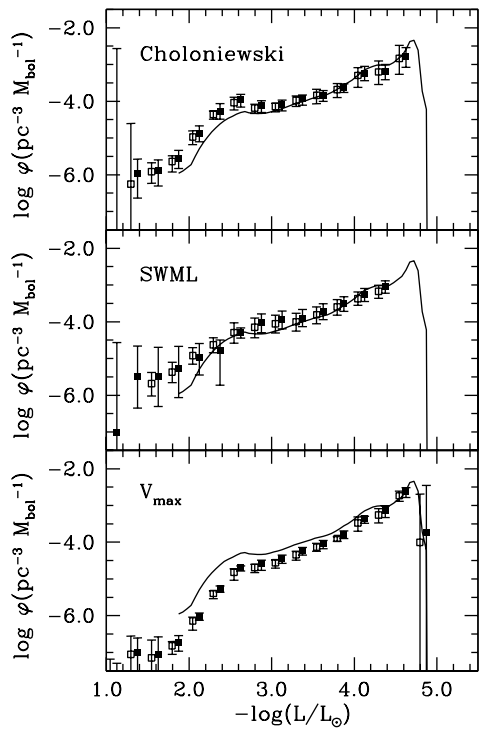}
\caption{Same  as Fig.~10 but  assuming now  an incompleteness  of the
         input  catalog  of  20\%  (filled  symbols)  and  40\%  (open
         symbols).  The open  symbols have  been moved  by $\Delta\log
         (L/L_{\sun})=-0.08$ for the sake of clarity.}
\end{figure}

Another  important concern  is how  the incompleteness  of  the sample
affects the  shape and  the location of  the cut--off of the retrieved
white  dwarf luminosity  function, and  how robust  are  the different
methods when  a sizeable  fraction of the  input sample  is discarded.
This is precisely the goal of  this section.  In order to assess these
effects we have first randomly  eliminated from the final input sample
discussed in  \S 4.6  (that is the  sample simulating the  white dwarf
luminosity function obtained  from the SDSS DR3) 20\%  and 40\% of the
white dwarfs  which pass all the selection  criteria, independently of
their magnitude,  proper motion, or tangential  velocity.  The results
are shown in Fig.~11.

As can be seen in the top panel of Fig.~11, the white dwarf luminosity
functions  obtained  using  the  Choloniewski  method  do  not  differ
considerably   from  those   previously   studied  in   \S  4.6   and,
consequently,  this  method   is  extremely  robust  against  possible
incompletenesses  of   the  input  sample,  even   under  the  radical
assumption that about 40\% of the white dwarfs in the input sample are
discarded  in the selection  process or,  simply, missed  whatever the
cause could  be.  For  the case in  which the  SWML method is  used we
stress that this method has the shortcomings already commented before:
firstly, it only recovers the shape of the luminosity function but not
the total density of white  dwarfs, and, secondly, it misses the faint
end of  the white dwarf  luminosity function.  However, there  are not
big differences in  the recovered shape of the  white dwarf luminosity
function, even for incompletenesses of  the order of 40\%. This is not
the case of the $1/\mathcal{V}_{\rm max}$ method which, as can be seen
in the  bottom panel of Fig.~11, largely  underestimates the resulting
white  dwarf  density  for  almost  all the  luminosity  bins.   Note,
however, that in this case the luminosity of the cut--off is correctly
retrieved,  independently of the  adopted incompleteness.   Hence, our
results show that the Choloniewski method is much more stable than the
$1/\mathcal{V}_{\rm max}$ method, even under extreme assumptions about
the completeness  of the  input sample used  to build the  white dwarf
luminosity  function.   On  the  other  hand,  for  the  case  of  the
$1/\mathcal{V}_{\rm max}$ method the size of the error bars is more or
less the same than in the  case in which the sample was complete. This
is  not  the   case  for  the  brightest  luminosity   bins  when  the
Choloniewski method is used.

\begin{figure}
\vspace{8cm}
\includegraphics{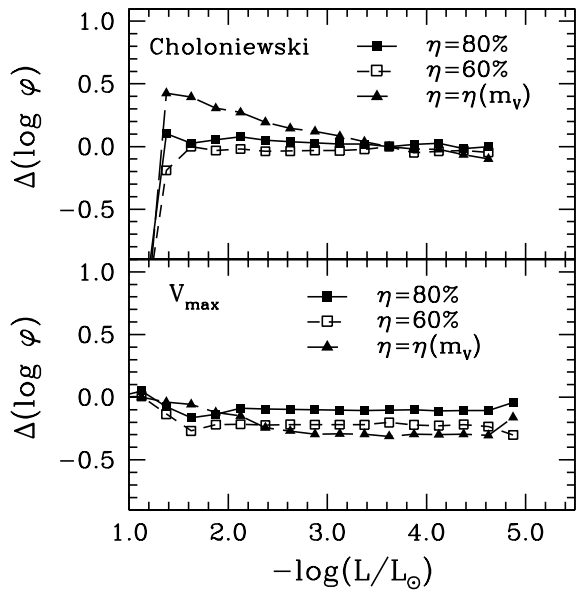}
\caption{Differences of the resulting white dwarf luminosity function,
         $\Delta  \log  \phi$, when  incompletenesses  of 20\%  (solid
         squares),  40\%  (open  squares)  and  a  linealy  decreasing
         completeness (solid  triangles) are assumed,  with respect to
         the white  dwarf luminosity  function obtained when  the full
         input sample is used. See text for details.}
\end{figure}

In a  second step  we have adopted  a different strategy.   Instead of
discarding  a given  percentage of  white dwarfs  regardless  of their
properties  we have  assumed than  the  input sample  is complete  for
apparent  magnitudes  $m_{\rm   V}=15.0$  and  that  the  completeness
decreases lineraly  to 60\% for  $m_{\rm V}=19.5$.  Fig.~12  shows the
results  of applying  this procedure  to the  input sample.   For this
figure  we  have  preferred  to  show  the  differences  $\Delta  \log
\varphi=\log \varphi^\prime-\log \varphi$  of the resulting luminosity
function,  $\log  \varphi^\prime$, with  respect  to  the white  dwarf
luminosity  function  obtained  using  the full  input  sample,  $\log
\varphi$, in order to better  visualize the results. The solid squares
are the differences when a completeness of $\eta=80$\% is assumed, the
open squares  are the  data for  a completeness of  only 60\%  and the
triangles  represent the  results obtained  when a  linarly decreasing
completeness is  adopted.  Fig.~12 shows  that the $1/\mathcal{V}_{\rm
max}$ method  underestimates the  white dwarf luminosity  function for
the vast majority of the  luminosity bins for all three cases, whereas
the Choloniewski method is quite  robust and, except for the brightest
luminosity  bins, is  rather insensitive  to the  completeness  of the
input  sample. Hence,  and from  this point  of view  the Choloniewski
method is clearly superior to the $1/\mathcal{V}_{\rm max}$ method.

\section{Conclusions}

We have performed a study  of the statistical reliability of the white
dwarf luminosity  function using different  estimators.  These include
the  classical $1/\mathcal{V}_{\rm  max}$ method,  and  two parametric
maximum-likelihood estimators, namely  the Choloniewski method and the
SWML or  the STY method,  depending on the adopted  cooling sequences.
In a first stage, for all  three estimators the input sample was drawn
from a controlled sample for  which we adopted the most simple cooling
law (Mestel 1952) and very schematic galactic inputs. This was done in
order  to study the  real behavior  of the  estimators and  to isolate
their  respectives advantages and  drawbacks. Nevertheless,  for these
numerical experiments the  observational selection criteria were fully
taken into  account. We have  found that for  a small sample  size the
$1/\mathcal{V}_{\rm max}$  method provides a  poor characterization of
the bright  end of the white  dwarf luminosity function  if the sample
selection procedure is not  done carefully.  Specifically, this method
produces  an artificial  deficit of  white dwarfs  at  moderately high
luminosities  when the  sample does  not contain  intrinsically bright
white dwarfs located at relatively  large distances.  This is a direct
consequence  of  the scarcity  of  intrinsically  bright white  dwarfs
which,  in turn,  is  a  consequence of  the  very short  evolutionary
time-scales of  these white dwarfs.  We have,  furthermore, shown that
this  is  possibly  the  case  of  the  catalog  of  spectroscopically
identified white dwarfs  of McCook \& Sion (1999),  for which very few
intrinsically bright white dwarfs  are present. Moreover, we have also
demonstrated   that   for   a   sample   size  of   300   stars,   the
$1/\mathcal{V}_{\rm  max}$ method  overestimates the  position  of the
drop--off  of   the  white  dwarf  luminosity  function.   This  is  a
consequence of the small number  of objects in the input sample which,
in  turn, forces  a coarse  binning.   We have  further discussed  the
effect of  the adopted scale height law  and we have found  that for a
sample size  of 300 stars its  effect cannot be  disentangled from the
effects of the sample  selection procedure. Additionally, we have also
shown that  the observational  procedure to assign  error bars  is too
optimistic  for small sample  sizes, with  realistic error  bars being
typically 10 times larger for a typical sample size of 300 objects.

We have explored two alternatives, the STY method and the Choloniewski
method. Both methods have been  widely used to build galaxy luminosity
functions with  satisfactory results, and  we have found that  for the
case of small  sample sizes they perform considerably  better than the
$1/\mathcal{V}_{\rm  max}$ method,  even if  none of  the  two methods
takes  into account  the bias  of proper  motion selected  samples. In
particular, the  STY method performs  best at recovering the  slope of
the luminosity function whereas  the Choloniewski method recovers best
the position of the cut--off. However, the STY method does not provide
the  true density  of white  dwarfs, whereas  the  Choloniewski method
does. 

We  have also  applied  the  two maximum  likelihodd  methods and  the
$1/\mathcal{V}_{\rm  max}$ method  to  a sample  of  300 white  dwarfs
obtained using  realistic stellar and  galactic inputs. In  this case,
instead of using  the STY method the SWML method  was used, given that
the  slope of  the increasing  branch  of the  white dwarf  luminosity
function is no  longer constant. We have found  that all three methods
present large  variances for the  brightest luminosity bins,  that the
SWML  method and  the  $1/\mathcal{V}_{\rm max}$  method retrieve  the
correct  location of  the  cut--off and that  the Choloniewski  method
underestimates the  number of faint  white dwarfs, resulting in  a bad
characterization of the maximum and of the cut--off of the white dwarf
luminosity function.

Finally, we have also applied these  three methods to a sample of 2000
white dwarfs, which is representative  of the sample used to build the
white  dwarf luminosity  function from  the  SDSS DR3  (Harris et  al.
2006).   This input  sample  was obtained  using up--to--date  cooling
sequences, realistic galactic inputs  and an accurate sample selection
procedure, following very precisely the prescriptions used for drawing
the final sample of white dwarfs  of the SDSS DR3.  We have found that
the performances of the Choloniewski method and of the $1/\mathcal{V}_
{\rm max}$ method are very similar, providing with reasonable accuracy
both the detailed shape of the white dwarf luminosity function and the
location  of the  cut--off.  Consequently, in  principle both  methods
could be used  in a real case, yielding similar  results. On the other
hand, the  SWML method does not  recover neither the  correct shape of
the  luminosity  function  nor  the  position  of  the  cut--off  and,
consequently,  should not be  used for  a real  sample.  We  have also
demonstrated   that  the  effects   of  the   scale  height   law  are
non--negligible for both  the Choloniewski and the $1/\mathcal{V}_{\rm
max}$ method.  Particularly, this  last method understimates the white
dwarf  density   for  the  brightest  luminosity   bins,  whereas  the
Choloniewski method  overestimates it.  For this last  input sample we
have  also analyzed the  effects of  the incompleteness,  finding that
only   the   Choloniewski  method   is   robust   when  the   possible
incompleteness  of the sample  is taken  into account,  retrieving the
correct total density of white dwarfs even for severe incompletenesses
of  the input  sample.  In  particular, the  $1/\mathcal{V}_{\rm max}$
method  severely  underestimates the  total  number  density of  white
dwarfs  for  sample  sizes  of   the  order  of  2000  stars  when  an
incompleteness of 20\% is adopted, whilst the Choloniewski method does
not,   being    thus   much    more   robust   than    the   classical
$1/\mathcal{V}_{\rm  max}$ method.   However,  the $1/\mathcal{V}_{\rm
max}$ method nicely recovers the position of the cut--off of the white
dwarf  luminosity  function,  whereas  the  Choloniewski  method  does
not.  In summary,  when  the input  sample  has a  sizeable number  of
objects   a   combination   of   both   the   Choloniewski   and   the
$1/\mathcal{V}_{\rm max}$  method provides reliable  determinations of
the white dwarf luminosity  function.  Other estimators, like the SWML
method, are  not recommended whatsoever  given that, firstly,  they do
not provide the true density of white dwarfs but only the shape of the
luminosity  function and,  secondly, they  do not  have  a performance
better than the other two methods.

\vspace{0.5 cm}

\noindent {\sl Acknowledgements.}  Part  of this work was supported by
the MCYT grants AYA04--094--C03-01 and 02, by the European Union FEDER
funds, and by the AGAUR.

\end{document}